\documentclass[preprintnumbers,amsmath,amssymbm,prd]{revtex4}
\usepackage{epsfig}
\usepackage{graphicx}

\begin{document}
\title{The instability spectra of near-extremal Reissner-Nordstr\"om-de Sitter black holes}
\author{Shahar Hod}
\affiliation{The Ruppin Academic Center, Emeq Hefer 40250, Israel}
\affiliation{ }
\affiliation{The Hadassah Institute, Jerusalem 91010, Israel}
\date{\today}

\begin{abstract}
\ \ \ The linearized dynamics of charged massive scalar fields in
the near-extremal charged Reissner-Nordstr\"om-de Sitter (RNdS)
black-hole spacetime is studied analytically. Interestingly, it is
proved that the non-asymptotically flat charged black-hole-field
system is characterized by {\it unstable} (exponentially growing in
time) complex resonant modes. Using a WKB analysis in the eikonal
large-mass regime $M\mu\gg1$, we provide a remarkably compact
analytical formula for the quasinormal resonant spectrum
$\{\omega_n(M,Q,\Lambda,\mu,q)\}_{n=0}^{n=\infty}$ which
characterizes the unstable modes of the composed
RNdS-black-hole-charged-massive-scalar-field system [here
$\{M,Q,\Lambda\}$ are respectively the mass, electric charge, and
cosmological constant of the black-hole spacetime, and $\{\mu,q\}$
are the proper mass and charge coupling constant of the linearized
field].
\end{abstract}
\bigskip
\maketitle

\section{Introduction}

The intriguing superradiant amplification phenomenon allows charged
integer-spin fields to extract Coulomb energy and electric charge
from various types of charged black holes
\cite{Bek1,Herch,Hodch1,Lich,Hodch2}. As a consequence, a charged
bosonic cloud surrounding a central charged black hole may grow
exponentially over time if the extracted energy is not radiated fast
enough to spatial infinity. Thus, the superradiant amplification
mechanism imposes a non-trivial threat to the stability of charged
black-hole spacetimes.

Interestingly, it has been proved in \cite{Hodcs} that
asymptotically flat charged Reissner-Nordstr\"om black-hole
spacetimes are linearly stable to charged massive scalar field
perturbations. On the other hand, it has recently been demonstrated
numerically \cite{Zhu,Kono1} that non-asymptotically flat charged
Reissner-Nordstr\"om-de Sitter (RNdS) black-hole spacetimes may
become unstable to perturbations of charged scalar fields whose
proper frequencies lie in the bounded superradiant regime
\cite{Noteunit}
\begin{equation}\label{Eq1}
{{qQ}\over{r_{\text{c}}}}<\omega<{{qQ}\over{r_+}}\  ,
\end{equation}
where $q$ is the charge coupling constant of the scalar field,
$\{Q,r_+\}$ are respectively the electric charge and outer horizon
radius of the central black hole, and $r_{\text{c}}$ is the radius
of the cosmological horizon which characterizes the black-hole
spacetime.

The main goal of the present paper is to use {\it analytical}
techniques in order to explore the physical and mathematical
properties of the intriguing instability phenomenon observed in the
highly interesting numerical works \cite{Zhu,Kono1}. In particular,
below we shall explicitly prove that the instability spectrum which
characterizes the composed
charged-RNdS-black-hole-charged-massive-scalar-field system can be
determined analytically in the near-extremal
$(r_{\text{c}}-r_+)/r_+\ll1$ regime.

\section{Description of the system}

We shall analyze the quasinormal resonant modes that characterize
the dynamics of a charged massive scalar field $\Psi$ which is
linearly coupled to a charged Reissner-Nordstr\"om-de Sitter black
hole (see \cite{QNM1,QNM2,QNM3} for excellent reviews on the
interesting phenomena of black-hole quasinormal resonances). The
non-asymptotically flat curved black-hole spacetime is described by
the line element \cite{Chan,Kono1,Noteunit}
\begin{equation}\label{Eq2}
ds^2=-f(r)dt^2+{1\over{f(r)}}dr^2+r^2(d\theta^2+\sin^2\theta
d\phi^2)\  ,
\end{equation}
where the radially-dependent metric function is given by the compact
expression
\begin{equation}\label{Eq3}
f(r)=1-{{2M}\over{r}}+{{Q^2}\over{r^2}}-{{r^2}\over{L^2}}\  .
\end{equation}
Here $\{M,Q,\Lambda\equiv3/L^2\}$ are respectively the mass,
electric charge \cite{NoteQ}, and the cosmological constant
\cite{Notelam,Notecl} of the black-hole spacetime. The radii
$\{r_-,r_+,r_{\text{c}}\}$ of the inner (Cauchy) horizon, outer
(event) horizon, and cosmological horizon which characterize the
RNdS black-hole spacetime are determined by the algebraic equation
\begin{equation}\label{Eq4}
f(r_*)=0\ \ \ \ \text{with}\ \ \ \ *\in\{-,+,\text{c}\}\  .
\end{equation}

The linearized dynamics of a scalar field $\Psi$ of proper mass
$\mu$ and charge coupling constant $q$ \cite{Noteqm} in the RNdS
black-hole spacetime is governed by the Klein-Gordon wave equation
\cite{Kono1,Stro}
\begin{equation}\label{Eq5}
[(\nabla^\nu-iqA^\nu)(\nabla_{\nu}-iqA_{\nu})-\mu^2]\Psi=0\  ,
\end{equation}
where $A_{\nu}=-\delta_{\nu}^{0}{Q/r}$ is the electromagnetic
potential of the charged black-hole spacetime. It is convenient to
expand the scalar field eigenfunction $\Psi$ in the form
\begin{equation}\label{Eq6}
\Psi(t,r,\theta,\phi)=\int\sum_{lm}{{\psi_{lm}(r;\omega)}\over{r}}Y_{lm}(\theta)e^{im\phi}e^{-i\omega
t} d\omega\  ,
\end{equation}
where the integer parameters $\{l,m\}$ (which are characterized by
the inequality $l\geq|m|$) denote the spherical and azimuthal
harmonic indices of the charged massive scalar field eigen-modes
\cite{Noteom}. Substituting the scalar field decomposition
(\ref{Eq6}) into the Klein-Gordon wave equation (\ref{Eq5}), one
obtains the ordinary differential equation
\begin{equation}\label{Eq7}
{{d^2\psi}\over{dy^2}}+V\psi=0\
\end{equation}
for the radial part of the scalar eigenfunction, where the tortoise
radial coordinate $y$ is related to the areal coordinate $r$ by the
simple differential relation \cite{Notemap}
\begin{equation}\label{Eq8}
{{dy}\over{dr}}=f^{-1}(r)\  .
\end{equation}
The effective black-hole-field radial potential
$V=V(r;M,Q,\Lambda,\omega,q,\mu,l)$ in the Schr\"odinger-like
differential equation (\ref{Eq7}) is given by \cite{Kono1}
\begin{equation}\label{Eq9}
V(r)=\Big(\omega-{{qQ}\over{r}}\Big)^2-f(r)G(r)\  ,
\end{equation}
where
\begin{equation}\label{Eq10}
G(r)=\mu^2+{{l(l+1)}\over{r^2}}+{1\over r}{{df}\over{dr}}\ .
\end{equation}

The quasinormal resonant modes, which characterize the composed
RNdS-black-hole-charged-massive-scalar-field system, are determined
by the Schr\"odinger-like ordinary differential equation (\ref{Eq7})
with the physically motivated boundary conditions of purely ingoing
waves at the outer (event) horizon of the black hole and purely
outgoing waves at the cosmological horizon of the spacetime
\cite{Kono1}:
\begin{equation}\label{Eq11}
\psi \sim
\begin{cases}
e^{-i(\omega-qQ/r_+)y} & \text{\ for\ \ \ } r\rightarrow r_+\ \
(y\rightarrow -\infty)\ ; \\ e^{i(\omega-qQ/r_{\text{c}})y} & \text{
for\ \ \ } r\rightarrow r_{\text{c}}\ \ \ (y\rightarrow \infty)\  .
\end{cases}
\end{equation}
Below we shall analyze the characteristic quasinormal resonant
spectra $\{\omega_n(M,Q,\Lambda,q,\mu,l)\}_{n=0}^{n=\infty}$ of the
composed near-extremal-RNdS-black-hole-charged-massive-scalar-field
system. In particular, we shall consider {\it complex} resonant
frequencies of the form
\begin{equation}\label{Eq12}
\omega=\omega_{\text{R}}-i\omega_{\text{I}}\  .
\end{equation}
Taking cognizance of Eqs. (\ref{Eq6}) and (\ref{Eq12}), one realizes
that resonant black-hole-field eigen-frequencies with
$\omega_{\text{I}}<0$ correspond to unstable (exponentially growing
in time) charged field modes, whereas resonant eigen-frequencies
with $\omega_{\text{I}}>0$ correspond to stable (exponentially
decaying) field modes.

\section{The regime of near-extremal Reissner-Nordstr\"om-de
Sitter black-hole spacetimes}

Interestingly, as we shall explicitly show below, the characteristic
quasinormal resonant frequencies
$\{\omega_n(M,Q,\Lambda,q,\mu,l)\}_{n=0}^{n=\infty}$ of the composed
charged-RNdS-black-hole-charged-field system can be determined {\it
analytically} in the regime
\begin{equation}\label{Eq13}
{{r_{\text{c}}-r_+}\over{r_+-r_-}}\ll1\
\end{equation}
of near-extremal black holes. In particular, in the regime
(\ref{Eq13}) one finds the simple functional relations [see Eq.
(\ref{Eq8})] \cite{Carne}
\begin{equation}\label{Eq14}
r(y)={{r_{\text{c}}e^{2\kappa_+y}+r_+}\over{1+e^{2\kappa_+y}+r_+}}\
\end{equation}
and
\begin{equation}\label{Eq15}
f(y)={{(r_{\text{c}}-r_+)\kappa_+}\over{2\cosh^2(\kappa_+y)}}\  ,
\end{equation}
where
\begin{equation}\label{Eq16}
\kappa_+={1\over2}\Big({{df}\over{dr}}\Big)_{r=r_+}\
\end{equation}
is the characteristic surface gravity of the black-hole outer
(event) horizon which, in the near-extremal regime (\ref{Eq13}), is
given by the compact dimensionless relation \cite{Carne,Notenz}
\begin{equation}\label{Eq17}
\kappa_+r_+={{(r_{\text{c}}-r_+)[1-2(Q/r_+)^2]}\over{2r_+}}\ll1\  .
\end{equation}

\section{The quasinormal resonance spectra of the charged massive scalar
fields in the near-extremal charged RNdS black-hole spacetimes}

In the present section we shall explicitly show that the complex
resonant spectra which characterize the composed
near-extremal-RNdS-black-hole-charged-massive-scalar-field system
can be determined analytically, using standard WKB techniques
\cite{WKB1,WKB2,WKB3,Will}, in the dimensionless large-mass regime
\cite{Noteltp}
\begin{equation}\label{Eq18}
\text{max}\big\{\kappa_+r_+,l(l+1)\big\}\ll \mu r_+\  ,
\end{equation}
in which case the effective black-hole-field radial potential
(\ref{Eq9}) can be approximated by
\begin{equation}\label{Eq19}
V[r(y)]=\Big(\omega-{{qQ}\over{r}}\Big)^2-\mu^2f(r)\cdot\{1+O[(\mu
r_+)^{-2}]\}.
\end{equation}

As explicitly proved in \cite{WKB1,WKB2}, the WKB resonance
condition which determines, in the large-frequency regime, the
fundamental complex resonant frequencies of the Schr\"odinger-like
radial equation (\ref{Eq7}) is given by the differential relation
\begin{equation}\label{Eq20}
{{iV(y_0)}\over{\sqrt{2V^{(2)}(y_0)}}}=n+{1\over 2}\ \ \ \ ; \ \ \ \ n=0,1,2,...
\end{equation}
where $V^{(k)}(y_0)\equiv [d^{k}V/dy^{k}]_{y=y_0}$ and $y_0(r_0)$ is
the local extremum point  [with $V^{(1)}(y_0)=0$] of the effective
black-hole-field radial potential $V(y)$. Taking cognizance of Eqs.
(\ref{Eq8}) and (\ref{Eq19}), one finds that this extremum point is
characterized by the functional relation \cite{Noteoep,KZ,TS}
\begin{equation}\label{Eq21}
\omega-{{qQ}\over{r_0}}={{\mu^2r^2_0}\over{2qQ}}\cdot{{f^{(1)}(y_0)}\over{f(y_0)}}\  ,
\end{equation}
where $f^{(k)}(y_0)\equiv [d^{k}f/dy^k]_{y=y_0}$.

Substituting the effective black-hole-field radial potential
(\ref{Eq19}) into Eq. (\ref{Eq20}), one finds
\begin{equation}\label{Eq22}
{{\Big(\omega-{{qQ}\over{r_0}}\Big)^2-\mu^2f(y_0)}\over
{\sqrt{\Big[4\Big({{qQ}\over{r^2_0}}\Big)^2-8\Big(\omega-{{qQ}\over{r_0}}\Big){{qQ}\over{r^3_0}}\Big]f^2(y_0)+
4\Big(\omega-{{qQ}\over{r_0}}\Big){{qQ}\over{r^2_0}}f^{(1)}(y_0)-2\mu^2f^{(2)}(y_0)}}}
= -i\Big(n+{1\over2}\Big)\  .
\end{equation}
The WKB condition (\ref{Eq22}), supplemented by the extremum
relation (\ref{Eq21}), determine the characteristic quasinormal
resonant spectra of the composed
near-extremal-RNdS-black-hole-charged-massive-scalar-field system.
Interestingly, as we shall now show, this (rather cumbersome)
equation can be solved {\it analytically} for the fundamental
complex resonant frequencies of the system which, in the eikonal
large-mass regime (\ref{Eq18}), are characterized by the strong
inequality [see Eqs. (\ref{Eq30}) and (\ref{Eq32}) below]
\begin{equation}\label{Eq23}
\omega_{\text{R}}\gg |\omega_{\text{I}}|\  .
\end{equation}
In particular, the real and imaginary parts of the WKB resonance
condition (\ref{Eq22}) can be decoupled in the large-frequency
(large-mass) regime (\ref{Eq23}). The real part of the resonance
equation is given by
\begin{equation}\label{Eq24}
\Big(\omega_{\text{R}}-{{qQ}\over{r_0}}\Big)^2-\mu^2f(y_0)=0\  ,
\end{equation}
whereas, to leading order in the small dimensionless ratio
$|\omega_{\text{I}}|/\omega_{\text{R}}$ [see (\ref{Eq23})], the
imaginary part of the resonance equation (\ref{Eq22}) is given by
\begin{equation}\label{Eq25}
2\Big(\omega_{\text{R}}-{{qQ}\over{r_0}}\Big)\omega_{\text{I}}=
\sqrt{\Big[4\Big({{qQ}\over{r^2_0}}\Big)^2-8\Big(\omega_{\text{R}}
-{{qQ}\over{r_0}}\Big){{qQ}\over{r^3_0}}\Big]f^2(y_0)+
4\Big(\omega_{\text{R}}-{{qQ}\over{r_0}}\Big)
{{qQ}\over{r^2_0}}f^{(1)}(y_0)-2\mu^2f^{(2)}(y_0)}\cdot\Big(n+{1\over2}\Big)\
.
\end{equation}

Substituting Eq. (\ref{Eq21}) into Eq. (\ref{Eq24}), one finds the
relation \cite{Noter0rp}
\begin{equation}\label{Eq26}
{{f^{(1)}(y_0)}\over{f^{3/2}(y_0)}}=\pm{{2qQ}\over{\mu
r^2_+}}\cdot[1+O(\kappa_+r_+)]\
\end{equation}
which, taking cognizance of the radial metric function (\ref{Eq15}) that
characterizes the near-extremal charged RNdS black-hole
spacetimes, yields the compact functional expression
\begin{equation}\label{Eq27}
\sinh(\kappa_+y^{\pm}_0)=\pm{{qQ}\over{\mu r_+}}[1-2(Q/r_+)^2]^{-1/2}\cdot[1+O(\kappa_+r_+)]\
\end{equation}
for the locations of the extremum points
$y^{\pm}_0=y^{\pm}_0(r^{\pm}_0)$ which characterize the effective
radial potential (\ref{Eq19}) of the composed black-hole-field
system \cite{Noterw}. From Eqs. (\ref{Eq14}) and (\ref{Eq17}) one
finds the dimensionless ratio
\begin{equation}\label{Eq28}
{{r(y)}\over{r_+}}=1+{{2\kappa_+r_+}\over{\big[1-2(Q/r_+)^2\big]
\big(1+e^{-2\kappa_+y}\big)}}\
\end{equation}
which, taking cognizance of Eq. (\ref{Eq27}), yields the relation
\cite{Notery}
\begin{equation}\label{Eq29}
{{r^{\pm}_0}\over{r_+}}=1+{{\kappa_+r_+}\over{1-2(Q/r_+)^2}}\cdot
\Bigg\{1\pm{{1}\over{\sqrt{1+\big({{\mu
r_+}\over{qQ}}\big)^2\big[1-2(Q/r_+)^2\big]}}}\Bigg\}+
O(\kappa^2_+r^2_+)\  .
\end{equation}
Substituting Eqs. (\ref{Eq15}), (\ref{Eq27}), and (\ref{Eq29}) into
Eq. (\ref{Eq21}), one obtains the expression
\begin{equation}\label{Eq30}
\omega^{\pm}_{\text{R}}={{qQ}\over{r_+}}-{{qQ}\over{1-2(Q/r_+)^2}}
\Bigg\{1\pm\sqrt{1+\Big({{\mu
r_+}\over{qQ}}\Big)^2\big[1-2(Q/r_+)^2\big]}\Bigg\}\cdot\kappa_+ +
O(\kappa^2_+r_+)\
\end{equation}
for the real parts of the resonant frequencies which characterize
the composed
near-extremal-RNdS-black-hole-charged-massive-scalar-field system.
It is worth pointing out that the resonant frequencies (\ref{Eq30})
are characterized by the inequalities [see Eq. (\ref{Eq17})]
\cite{Notetk}
\begin{equation}\label{Eq31}
\omega^+_{\text{R}}<{{qQ}\over{r_{\text{c}}}}\ \ \ \ \text{and}\ \ \
\ \omega^-_{\text{R}}>{{qQ}\over{r_+}}\  .
\end{equation}
Taking cognizance of the relations (\ref{Eq1}) and (\ref{Eq31}), one
deduces that these resonant frequencies lie outside the superradiant
regime of the system.

Substituting Eqs. (\ref{Eq8}), (\ref{Eq15}), (\ref{Eq27}), and
(\ref{Eq30}) into Eq. (\ref{Eq25}), one obtains the remarkably
compact functional relation
\begin{equation}\label{Eq32}
\omega^{\pm}_{\text{I}}=\mp\kappa_+\cdot\big(n+{1\over2}\big)\cdot[1+O(\kappa_+r_+)]\
\end{equation}
for the imaginary parts of the quasinormal resonant frequencies
which, in the eikonal large-mass regime (\ref{Eq18}), characterize
the linearized dynamics of the charged massive scalar fields in the
near-extremal ($\kappa r_+\ll1$) charged RNdS black-hole spacetimes.
Taking cognizance of Eqs. (\ref{Eq23}), (\ref{Eq30}), and
(\ref{Eq32}), one learns that our analysis is valid in the
dimensionless regime \cite{Notensl}
\begin{equation}\label{Eq33}
\kappa_+r_+\cdot(n+{1\over2})\ll qQ\  .
\end{equation}

\section{The regime of validity of the WKB approximation}

It is important to stress the fact that the WKB functional
expressions (\ref{Eq30}) and (\ref{Eq32}) for the real and imaginary
parts of the complex resonant frequencies which characterize the
composed near-extremal-RNdS-black-hole-charged-massive-scalar-field
system are valid under the assumption that higher-order correction
terms that appear in the WKB resonance condition (\ref{Eq20}) can be
neglected. In particular, it was proved in
\cite{WKB1,WKB2,WKB3,Will} that, taking into account higher-order
spatial derivatives of the effective radial potential $V(y)$ [see
Eq. (\ref{Eq19})], one obtains the higher-order correction term
\begin{equation}\label{Eq34}
\Lambda(n)={{1}\over{\sqrt{2V^{(2)}_0}}}\Bigg[-
{{7+15(2n+1)^2}\over{288}}\cdot\Big({{V^{(3)}_0}\over{V^{(2)}_0}}\Big)^2+
{{1+(2n+1)^2}\over{32}}\cdot{{V^{(4)}_0}\over{V^{(2)}_0}}\Bigg]\
\end{equation}
on the r.h.s of the black-hole-field resonance condition
(\ref{Eq20}). Thus, our analytically derived results (\ref{Eq30})
and (\ref{Eq32}) are valid provided [see Eqs. (\ref{Eq20}) and
(\ref{Eq34})] \cite{Notewkcon}
\begin{equation}\label{Eq35}
{{\Lambda(n)}\over{n+{1\over 2}}}\ll1\  .
\end{equation}

Using Eqs. (\ref{Eq8}), (\ref{Eq15}), (\ref{Eq19}), and
(\ref{Eq27}), one finds the expression
\begin{equation}\label{Eq36}
\Lambda(n)=-{{4(2n+1)^2\big({{qQ}\over{\mu
r_+}}\big)^2+[1+(2n+1)^2][1-2(Q/r_+)^2]}\over{8\sqrt{(qQ)^2+(\mu
r_+)^2[1-2(Q/r_+)^2]}}}\
\end{equation}
for the higher-order correction term (\ref{Eq34}) in the WKB
approximation. Taking cognizance of (\ref{Eq35}) and (\ref{Eq36}),
one deduces that the analytical expressions (\ref{Eq30}) and
(\ref{Eq32}) for the real and imaginary parts of the composed
black-hole-field quasinormal resonant frequencies, which have been
derived using the WKB resonance condition (\ref{Eq20}), are valid in
the dimensionless physical regime
\begin{equation}\label{Eq37}
{{\max\big\{\big({{qQ}\over{\mu
r_+}}\big)^2,1-2(Q/r_+)^2\big\}}\over{\sqrt{(qQ)^2+(\mu
r_+)^2[1-2(Q/r_+)^2]}}}\ll(n+{1\over2})^{-1}\  .
\end{equation}
Interestingly, one learns from (\ref{Eq37}) that, for $\mu r_+\gg
n+1/2$ \cite{Notemrb}, the instability of the near-extremal charged
RNdS black-hole spacetimes can occur for arbitrarily small [with the
constraint (\ref{Eq33}) \cite{Notensl}] values of the dimensionless
charge coupling constant $qQ$. In addition, one deduces from
(\ref{Eq37}) that the physical parameter $qQ$ is bounded from above
by the simple dimensionless relation
\begin{equation}\label{Eq38}
qQ\ll (\mu r_+)^2\cdot(n+{1\over2})^{-1}\  .
\end{equation}

\section{Summary}

The Reissner-Nordstr\"om-de Sitter black-hole spacetimes describe a
family of charged solutions to the coupled Einstein-Maxwell field
equations in non-asymptotically flat de Sitter spacetimes. It has
recently been demonstrated numerically in the very interesting works
\cite{Zhu,Kono1} that, as opposed to the asymptotically flat
Reissner-Nordstr\"om black-hole spacetimes \cite{Hodcs}, the
non-asymptotically flat Reissner-Nordstr\"om-de Sitter black-hole
spacetimes may become unstable to perturbations of charged scalar
fields in the bounded superradiant regime (\ref{Eq1}).

In the present paper we have studied {\it analytically} the
instability spectrum which characterizes the linearized dynamics of
charged massive scalar fields in the charged near-extremal (with
$\kappa_+r_+\ll1$) Reissner-Nordstr\"om-de Sitter black-hole
spacetimes. In particular, using WKB techniques in the dimensionless
physical regime [see Eqs. (\ref{Eq33}), (\ref{Eq38}) and
\cite{Notensl,Notemrb}]
\begin{equation}\label{Eq39}
n+{1\over2}\ll\mu r_+\ \ \ \ \text{with}\ \ \ \
\kappa_+ r_+\cdot(n+{1\over2})\ll qQ\ll (\mu r_+)^2\cdot(n+{1\over2})^{-1}\  ,
\end{equation}
we have derived the remarkably compact analytical formula [see Eqs.
(\ref{Eq30}) and (\ref{Eq32})]
\begin{equation}\label{Eq40}
\omega_{\text{R}}={{qQ}\over{r_+}}-{{qQ}\over{1-2(Q/r_+)^2}}
\Bigg\{1+\sqrt{1+\Big({{\mu
r_+}\over{qQ}}\Big)^2\big[1-2(Q/r_+)^2\big]}\Bigg\}\cdot\kappa_+
-i\kappa_+\cdot\big(n+{1\over2}\big)+O(\kappa^2_+r_+)
\end{equation}
for the complex resonant frequencies which characterize the
instability spectrum of the composed
charged-near-extremal-RNdS-black-hole-charged-massive-scalar-field
system.

\bigskip
\noindent
{\bf ACKNOWLEDGMENTS}
\bigskip

This research is supported by the Carmel Science Foundation. I would
like to thank Yael Oren, Arbel M. Ongo, Ayelet B. Lata, and Alona B.
Tea for helpful discussions.



\begin{thebibliography}{99}
\bibitem{Noteunit} We shall use natural units in which $G=c=\hbar=k_{\text{B}}=1$.
\bibitem{NoteQ} We shall henceforth assume $Q>0$ for the black-hole charge without loss of generality.
\bibitem{Notelam} Note that Reissner-Nordstr\"om-de Sitter
black-hole spacetimes are characterized by the inequality
$\Lambda>0$.
\bibitem{Notecl} Here $L$ is the characteristic curvature radius of the
asymptotically de Sitter spacetime.
\bibitem{Noteqm} Note that the field parameters $\mu$ and $q$ stand for
$\mu/\hbar$ and $q/\hbar$, respectively. These physical parameters therefore
have the dimensions of $($length$)^{-1}$.
\bibitem{Noteom} We shall henceforth omit the harmonic indices $\{l,m\}$ for brevity.
\bibitem{Notemap} Note that the differential relation (\ref{Eq8})
maps the radial black-hole interval $r\in(r_+,r_{\text{c}})$ to the
interval $y\in(-\infty,\infty)$.
\bibitem{Notenz} Note that RNdS black-hole spacetimes are
characterized by the relation $1-2(Q/r_+)^2\geq0$ \cite{Carne}, and
thus $\kappa_+r_+\geq0$.
\bibitem{Noteltp} Note that $r\cdot df/dr=O(\kappa_+ r_+)\ll1$ for
the near-extremal RNdS black-hole spacetimes [see Eqs. (\ref{Eq8}),
(\ref{Eq15}), and (\ref{Eq17})].
\bibitem{Noteoep} It is worth mentioning that Konoplya and Zhidenko \cite{KZ}
have recently shown in a very interesting work that, in
asymptotically flat spacetimes, the WKB approximation may fail for
massive fields. This observation stems from the fact that, for
asymptotically flat spacetimes, the effective black-hole-field
potential may be characterized by the presence of a local minimum
far from black hole (where the mass term dominates the effective
potential of the system). It should be emphasized, however, that in
the present paper we consider a different physical scenario -- a
non-asymptotically flat (RNdS) black-hole spacetime, in which case
the mass term (and also the centrifugal term) in the effective
potential [see Eqs. (\ref{Eq3}), (\ref{Eq9}), and (\ref{Eq10})]
vanishes near the cosmological horizon. Thus, in our
non-asymptotically flat black-hole spacetime, the
frequency-dependent radial potential (\ref{Eq19}) of the composed
black-hole-charged-field system is characterized in the large-mass
regime (\ref{Eq18}) by only one extremum (maximum) point. It is
worth mentioning that a similar conclusion, regarding the
barrier-like spatial form of the effective potential for
black-hole-massive-fields systems in the non-asymptotically flat
Schwarzschild-de-Sitter black-hole spacetime, has recently been
reached in \cite{TS}.
\bibitem{Noter0rp} Here we have used the relation $r_0=r_+[1+O(\kappa_+r_+)]$ for the
near-extremal black holes in the dimensionless regime (\ref{Eq13}).
\bibitem{Noterw} It is important to emphasize the fact that the
effective black-hole-field radial potential (\ref{Eq19}) is
frequency-dependent. Thus, for a given value of the resonant
frequency $\omega_{\text{R}}$ [see Eq. (\ref{Eq30}) below], the
frequency-dependent radial potential (\ref{Eq19}) is characterized
by only one local extremum point which, as we shall explicitly prove
below, is characterized by the relation $V^{(2)}(y_0)>0$.
\bibitem{Notery} Here we have used the identities $2/(1+e^{-2\alpha})=1+\tanh(\alpha)$ and
$\cosh(\alpha)=\sqrt{1+\sinh^2(\alpha)}$.
\bibitem{Notetk} I would like to thank Professor Ran Li for pointing
out the fact that the resonant frequency $\omega^+_{\text{R}}$ is
characterized by the inequality
$\omega^+_{\text{R}}<{{qQ}\over{r_{\text{c}}}}$.
\bibitem{Notensl} It is worth emphasizing again that in the present paper
we consider the physical regime of near-extremal RNdS black-hole
spacetimes with $\kappa_+r_+\ll1$. Thus, the inequality (\ref{Eq33})
does not necessarily imply that the dimensionless charge coupling
constant $qQ$ is large [see Eq. (\ref{Eq38}) below].
\bibitem{Notewkcon} It is important to emphasize that the convergence of the WKB series
is guaranteed only asymptotically and thus the relation (\ref{Eq35})
may serve only as a supportive indication (not a rigorous
mathematical proof) for the validity of the WKB treatment presented
in this work.
\bibitem{Notemrb} It is important to point out that the strong inequality (\ref{Eq37})
implies the dimensionless relation $\mu r_+\gg n+1/2$. To see this,
we note that the relation $\sqrt{(qQ)^2+(\mu
r_+)^2[1-2(Q/r_+)^2]}\gg n+1/2$ provides a necessary condition for
the validity of (\ref{Eq37}). This implies that $\mu r_+\gg n+1/2$
and/or $qQ\gg n+1/2$. Let us assume that $\mu r_+$ is not large [and
therefore $qQ$ must be large with $qQ\gg \text{max}\{n+1/2,\mu
r_+\}$]. In this case, one deduces from (\ref{Eq37}) the strong
inequalities $\text{max}\{n+1/2,\mu r_+\}\ll qQ\ll (\mu
r_+)^2\cdot(n+{1\over2})^{-1}$, which immediately imply the
dimensionless relation $\mu r_+\gg n+1/2$.
\end{thebibliography}
\end{document}